\documentclass[
  twoside,
  twocolumn,
  aps,
  10pt,
  prb,
  preprintnumbers,
  floatfix,
  showpacs,
  citeautoscript,
]{revtex4-1}

\usepackage{amsmath}
\usepackage{amssymb}
\usepackage{color}
\usepackage{graphicx}
\usepackage{tabularx}
\usepackage{subfigure}
\usepackage{dcolumn}
\usepackage{times}

\newcommand{\phys}{
  Chalmers University of Technology,
  Department of Physics,
  Gothenburg, Sweden
}
\newcommand{\mctwo}{
  Chalmers University of Technology,
  Department of Microtechnology and Nanoscience, MC2,
  Gothenburg, Sweden
}

\newcommand{\sect}[1]{Sect.~\ref{#1}}
\newcommand{\fig}[1]{Fig.~\ref{#1}}
\newcommand{\eq}[1]{Eq.~(\ref{#1})}
\newcommand{\tab}[1]{Table~\ref{#1}}

\renewcommand{\vec}[1]{\ensuremath\boldsymbol{#1}}
\renewcommand{\epsilon}[0]{\varepsilon}
\newcommand{\rr}[0]{\vec{r}}

\begin{document}

\preprint{}
\pacs{}

\title{
  Finite-temperature properties of non-magnetic transition metals: \\
  Comparison of the performance of constraint-based semi and nonlocal functionals
}

\author{Leili Gharaee}
\author{Paul Erhart}
\email{erhart@chalmers.se}
\affiliation{\phys}

\author{Per Hyldgaard}
\email{hyldgaar@chalmers.se}
\affiliation{\mctwo}

\begin{abstract}
We assess the performance of nonempirical, truly nonlocal and semi-local
functionals with regard to structural and thermal properties of $3d$, $4d$,
and $5d$ non-magnetic transition metals. We focus on constraint- based
functionals and consider the new consistent-exchange van der Waals density
functional version vdW-DF-cx [Phys. Rev. B \textbf{89}, 035412 (2014)], the
semi-local PBE [Phys. Rev. Lett. \textbf{77}, 3865 (1996)] and PBEsol
functionals [Phys. Rev. Lett. \textbf{100}, 136406 (2008)] as well as the AM05
meta-functional [Phys. Rev. B \textbf{72}, 085108 (2005)]. Using the quasi-
harmonic approximation structural parameters, elastic response, and thermal
expansion at finite temperatures are computed and compared to experimental
data. We also compute cohesive energies explicitly including zero-point
vibrations. It is shown that overall vdW-DF-cx provides an accurate
description of thermal properties and retains a level of transferability and
accuracy that is comparable to or better than some of the best constraint-
based semi-local functionals. Especially, with regard to the cohesive energies
the consistent inclusion of spin polarization effects in the atoms turns out
to be crucial and it is important to use the rigorous spin- vdW-DF-cx
formulation [Phys. Rev. Lett. \textbf{115}, 136402 (2015)]. This demonstrates
that vdW-DF-cx has general-purpose character and can be used to study systems
that have both sparse and dense electron distributions.
\end{abstract}

\maketitle


\section{Introduction}

The adiabatic connection formula (ACF) enables a formal determination of all
exchange-correlation (XC) effects in density functional theory (DFT)
\cite{HarJon74, GunLun76, LanPer77}. The XC energy density functional (DF)
$E_{xc}$ can be seen as the electrostatic binding of electrons with its
associated XC hole. Good $E_{xc}$ approximations represent the core of DFT and
$E_{xc}$ formulations reflect insight pertaining to the collective response of
the interacting electron gas \cite{GunLun76, LanPer77}. Important progress
followed by enforcing hole conservation and other physical constraints in the
formulation of the local density approximation (LDA) as well as semi-local
functionals based on the generalized gradient approximation (GGA). Hole
conservation underpins, for example, the PBE functional \cite{PBE96}, which
has proven to be highly successful as a general-purpose functional for
problems where the physical behavior is governed by the response at large
electron concentrations. Specifically, the PBE functional is accurate for both
hard materials and individual molecules and has accordingly found widespread
applications \cite{BurkeDFTperspectives}.

The last decade has witnessed the successful introduction of the van der Waals
(vdW) density functional (vdW-DF) method \cite{RydDioJac03, DioRydSch04,
ThoCooLi07, BerJCP14, BerIOP15}, launched as a systematic extension
\cite{RapAsh91, AndLanLun96, HylBerSch14} of LDA and GGA. Unlike local and
semi-local functionals such as the latter two it describes also the much
larger class of sparse systems \cite{LanLunCha09}, e.g., molecular solids,
layered materials, and weak chemisorption cases, in which binding across
internal voids arises from a truly nonlocal, vdW-type binding. The Chalmers-
Rutgers vdW-DF method is focused on the electron response and has both regular
releases (vdW-DF1 \cite{DioRydSch04}, vdW-DF2 \cite{LeeMurKon10}, and
vdW-DF-cx \cite{BerHyl14}) and variants (including vdW-DF-C09 \cite{Coo09}, optB88,
optPBE \cite{KliBowMic10}, optB86b,\cite{KliBowMic11} and rev-vdW-DF2
\cite{Ham14}). It has recently been extended with a rigorous spin formulation
that reflects the vdW-DF design logic \cite{ThoZulArt15}. There are also
related but alternative formulations of the nonlocal-correlation term in the
Vydrov-van Voorhis family (VV09 \cite{VV09}, vdW-DF-09 \cite{vdW-DF-09} and
VV10 \cite{VV10}) and approaches that emphasize multipole response and mutual
coupling of exhange holes \cite{Becke07p154108, Becke05p154101,
Becke05p154104}. In addition, there are approaches that focus on the dipole
and multipole response of orbitals, atoms, and clusters (typically obtained
outside of the ground-state DFT framework), compute the mutual coupling energy
and add it to a traditional ground state DFT description \cite{Gri06, TS09,
Silvestrelli08p53002, Silvestrelli09p5224, GriAntEhr10, Ambrosetti12p73101,
TS-MBD, TSvdWsurf, RuzPerTao12, TaoPer14, TaoRap14, AmbSil16}. It is commonly
required that the chosen approach must be accurate both when there are
important regions of low-electron concentration, i.e., sparse matter such as
in intermolecular binding \cite{LanLunCha09}, and when studying dense matter,
i.e., harder materials such as transition metals.

The recent consistent-exchange version, vdW-DF-cx \cite{BerHyl14,   BerJCP14,
HylBerSch14}, uses the vdW-DF plasmon-pole response description also to
determine the \emph{semilocal} component of the vdW-DF method. vdW-DF-cx is
determined by a dielectric-response description that automatically enforces
current conservation in the screening response \cite{Dion,HylBerSch14}. The
vdW-DF-cx method effectively uses the same plasmon-pole response to define
both gradient-corrected exchange and nonlocal correlations. This new vdW-DF
version thereby minimizes a hidden cross-over term $\delta E_x^0$ that
generally enters in the vdW-DF family of functionals \cite{BerHyl14}.

vdW-DF-cx has already proven itself accurate and useful in a number of
problems that involve both regions of sparse and dense electron distributions
such as molecular dimers \cite{S22, BerHyl14}, layered materials
\cite{BerJCP14, torbjorn14, ErhHylLin15, LinErh16,   MurSunIma15,
SadSanLam15}, semiconductors \cite{BerJCP14,MehDorZhu16}, molecular crystals
\cite{RanPRB16,   BrownAltvPRB16}, adsorption processes \cite{ThoZulArt15,
LofGroMot16} as well as weak chemisorption, molecular switching, and molecular
self assembly \cite{ThoZulArt15,Arnau16, LofGroMot16}.  The ACF foundation and
the emphasis on conservation laws in the vdW-DF-cx construction further
suggests a general-purpose nature\cite{BerJCP14} and motivates a comprehensive
investigation of its performance also for regular dense matter. This is
particularly interesting since earlier members of the vdW-DF family have
repeatedly been found to yield an inferior description of traditional bulk
materials, in particular the late transition metals \cite{LofGroMot16}.

Here, we benchmark vdW-DF-cx for thermo-physical properties of non-magnetic
transition metals, for which extensive experimental data is available for
comparison \cite{KittelRefData,   ThermalExpandData}. Specifically, we
consider lattice parameters, thermal expansion, and bulk moduli at finite
temperature as well as the cohesive energies including zero-point
contributions at the level of he quasi harmonic approximation (QHA). This data
set explicitly tests not only the description of energy and structure but also
forces, going beyond the set of properties commonly considered in comparative
assessments of XC functionals. In addition to vdW-DF-cx we consider the
constraint-based semi-local functionals, PBE \cite{PBE96}, PBEsol
\cite{PBEsol08} and AM05 \cite{AM05}.

We show that vdW-DF-cx meets and exceeds the performance of the PBE, PBEsol
and AM05 functionals for nonmagnetic transition metals. This suggests that
vdW-DF-cx provides a good, balanced description of nonlocal exchange and
nonlocal correlation also in this type of materials. vdW-DF-cx thus remains a
candidate for serving as a general-purpose materials-theory tool, working for
both hard and soft matter \cite{BerJCP14}.

The remainder of this paper is organized as follows. The next section provides
an overview of the constraint-based functionals considered in the present work
while methodological aspects are compiled in \sect{sect:computational-details}.
Section~\ref{sect:results} describes the main results before
\sect{sect:summary} provides a summary and conclusions. A detailed compilation
of results including a per-element comparison with experimental data can be
found in the Supplementary Material \cite{SuppMatCite}.

\section{Constraint-based nonlocal functionals}
\label{sect:constraint-based-nonlocal-functionals}

\subsection{General aspects}

Comparisons among constraint-based nonlocal functionals are valuable in our
drive to further improve truly nonlocal DFs. Several previous studies have
shown that some vdW-DF approaches can work well for solids \cite{ZiaKleSch07,
KliBowMic10, KliBowMic11,   BerJCP14,Ham14}. While this has helped build trust
in the vdW-DF method, constraint-based functionals such as vdW-DF-cx, PBE,
PBEsol, and AM05 are all linked to the ACF and conservation \cite{PBE96,
PBEsol, AM05, BerHyl14, HylBerSch14} and can thus be expected to yield good
transferability. Yet as briefly reviewed below different physical aspects were
emphasized in their construction and one can thus expect to gain insight into
strengths and limitations of each approach exactly because these functionals
are each representatives of a specific design logic. By focusing the present
benchmark on constraint-based functionals we are thus able to draw more
general conclusions.

The four constraint-based functionals considered in the present work share
some common traits while also having some distinct differences in their design
logic, making it interesting to contrast their performance. All matter has
internal surfaces with a variation between higher and lower electron density
regions and insight from surface physics underpins all designs. It led
Langreth and Perdew to the early GGA \cite{LanPer77} and it entered the
specification of gradient-corrected correlation in PBEsol \cite{PBEsol08} and
AM05 \cite{AM05}. These concepts are also central to the development of the
vdW-DF method \cite{RydDioJac03, BerIOP15}, which takes the surface idea,
however, further than in the GGA, noting that a semi-local representation of
the electron-gas response does not retain a full description of the
electrodynamical coupling among (GGA) XC holes \cite{RapAsh91, AndLanLun96,
RydDioJac03, HylBerSch14, BerIOP15}. The electrodynamical coupling is
relevant, for example, when there are multiple interacting density fragments
(molecules or surfaces) separated by a region with low electron concentration
\cite{Rydberg03p126402, HylBerSch14}.

The Fermi wave vector $k_F(\rr)=\left[3\pi^2 n(\rr)\right]^{1/3}$ sets a local
energy scale via the LDA exchange energy per particle,
$\varepsilon_{x}^\text{LDA}(\rr)=-(3/4\pi) k_F(\rr)$. We take semi-local
functionals to imply that the XC hole form or the energy per particle
exclusively depend on the local (spin) density $n(\rr)$ and the local scaled
gradient $s(\rr)=|\nabla n|/\left[2 n(\rr)   k_F(\rr)\right]$. Semi-local GGA
functionals can be expressed via the local variation in the XC energy per
particle \begin{align}   \varepsilon_{xc}^\text{GGA}(\rr) \equiv
F_{xc}\left(n(\rr),s(\rr)\right) \varepsilon_{x}^\text{LDA}(\rr).
\label{eq:enhanceXCf} \end{align} Here, the XC enhancement factor
$F_{xc}(n(\rr),s(\rr))$ reflects the physical nature of the associated semi-
local XC hole \cite{PBE96,   PBEsol08}.

\subsection{The PBE functional}

The PBE functional \cite{PBE96} is an important example of a constraint-based
GGA \cite{BurkeDFTperspectives,   BeckeDFTperspectives}. The PBE was designed
by first constructing a numerical GGA with an enhancement factor
$\tilde{F}_{xc}$ that reflects conditions on the shape of the semi-local XC
hole description and, in a subsequent step, by extracting an analytical form
for the PBE $F_{xc}$ for practical use. One can expect a high degree of
transferability because it is anchored in conservation laws \cite{LanPer77,
GunLun76}. In fact the PBE functional has had a huge impact on materials
theory and has turned out to be an extremely successful general-purpose
functional for systems with dense electron distributions including both
individual molecules and hard materials \cite{BurkeDFTperspectives}.

\subsection{The PBEsol functional}

One of the best performing constraint-based semi-local functionals for
condensed matter is the PBEsol functional \cite{PBEsol08}. While, as
in the case of PBE, the nature of screened many-body response and the
XC hole were emphasized during its construction, its authors also
relied on other formal results in its design. The GGA framework that
underpins both PBE and PBEsol is very powerful but it is not possible
to satisfy all constraints at the same time.
\footnote{
  The relation between PBE and PBEsol reflects in part the physical insight
used for picking the form of gradient-corrected exchange enhancements.
\cite{PBEsol08} Both functionals use conserving (albeit different)
approximations for the semi-local XC hole. In the   case of the PBE functional
one arrives at a small-$s$ expansion that   is also suggested by exact-scaling
results for atomic-like   high-density regions. \cite{PBEsol08,
BurkeDFTperspectives,     BeckeDFTperspectives} By contrast, in the
construction of PBEsol   one obtains a behavior consistent with diagrammatic
results for pure   exchange in the weakly perturbed electron gas.
}
The PBE functional is highly transferable and works very well for both
molecular formation energies and the structure and energies of hard
materials. As a result of the diagrammatic (gradient-expansion)
emphasis PBEsol yields an even better description of the structure of
hard materials \cite{PBEsol08}.

\subsection{The AM05 functional}

The AM05 constraint-based functional \cite{AM05} performs very well
for describing the structure of dense materials. It provides an, in
principle, exact account of exchange effects for surfaces, i.e., the
boundary between regions of higher and lower electron densities,
whereas for internal regions this surface-exchange description is
merged with that of the LDA. Like PBEsol, the AM05 functional extracts
the gradient-corrected correlation from a study of the jellium surface
energy. While it has different roots than PBE and PBEsol it is also
constraint-based and can be viewed as a semi-local functional since it
is possible to express the energy per particle variation using
\eq{eq:enhanceXCf}. Like the regular GGAs, it lacks an account of
truly nonlocal correlation effects.

\subsection{The vdW-DF framework}

The vdW-DF method \cite{AndLanLun96, RydDioJac03, DioRydSch04, ThoCooLi07,
BerHyl14, BerIOP15} represents a systematic nonempirical extension of both LDA
and the semi-local GGA description \cite{BerJCP14, HylBerSch14, ThoZulArt15}.
The very first version of this method was conceived two decades ago starting
from a simple Ashcroft picture \cite{RapAsh91} of vdW binding in the itinerant
electron gas \cite{HylBerSch14, BerIOP15}. It provides seamless integration
with a GGA-type description while enforcing conservation laws on the
underlying many-body response description \cite{DioRydSch04, ThoCooLi07,
HylBerSch14, BerIOP15}. The method predates the PBEsol and AM05 functionals
and its origin \cite{AndLanLun96} actually coincides with the launching of the
PBE functional \cite{PBE96}. The vdW-DF method captures vdW forces among
dimers in the asymptotic and the binding limits\cite{DioRydSch04,
Puzder06p164105, Li08p9031, Berland10p134705, Berland11p1800, CalHam15, rev6}
as well as attraction between two-dimensional layers \cite{Rydberg03p126402,
Langreth05p599, torbjorn14, ErhHylLin15, LinErh16}. Importantly, it also
captures the more general problem when nonlocal correlation forces compete
with other types of forces, \cite{BerJCP14} for example, giving rise to
binding across important regions of low electron densities \cite{LanLunCha09,
BerIOP15}. The method was expected to be relevant in first-principle DFT for
both pure vdW problems including regular physisorption \cite{LeeH2CuI,
LeeH2CuII, Londero12p424212, LofGroMot16}, porous-materials gas absorption
\cite{Yao12p064302, Tan12p3153, poloni_co2_2012, Liu12p3343, Lee14p698,
poloni_understanding_2014, TanZulGao15, ZulFulFer16, PolKim16, KuiHanLin16},
and DNA base-pair interactions \cite{OrScBe05, CooThoPuz07, LiCoThLuLa09,
Berland11p135001, Le12p424210}.  It was also quickly realized that truly
nonlocal correlations affect materials descriptions much more broadly than
what was perhaps originally anticipated \cite{ChaBorSch06, ZiaKleSch07,
Kleis08p205422, JohKleLun08, BerEinHyl09, LanLunCha09, rev6, rev8, BerIOP15}.
The recent vdW-DF formulations are, for example, proving themselves valuable
in the treatment of organic-inorganic interfaces and general weak-
chemisorption problems \cite{Chakarova-Kack06p155402, Chakarova-Kack06p146107,
KliBowMic10, Li12p121409, TSvdWsurf, ThoZulArt15, MauRuiTka15, Arnau16,
LofGroMot16, rev2}.

The vdW-DF framework formally constitutes an ACF recast
\cite{DioRydSch04, HylBerSch14, BerIOP15}
\begin{align} 
  E_{xc} = \int_0^\infty \, \frac{du}{2\pi}\, \text{Tr}\{\ln(\nabla
  \epsilon(iu) \cdot \nabla G)\} - E_\text{self}.
  \label{eq:vdWDFfram}
\end{align}
Here, $E_\text{self}$ and $G$ denote the Coulomb self-energy and Green
function, respectively, $u$ is a complex frequency, and $\epsilon$
denotes a suitable approximation for a scalar, nonlocal dielectric
function. The trace is over all spatial coordinates. The vdW-DF
framework has exact screening and it defines $\epsilon$ via a
plasmon-pole response description that reflects constraints like the
$F$ sum rule conservation \cite{DioRydSch04, BerJCP14, HylBerSch14, BerIOP15}.  The vdW-DF method begins with a plasmon-pole
approximation for screened response treated at the GGA level, i.e., by
choosing $\epsilon(iu)$ so that it reflects the shape of an internal
XC hole corresponding to a GGA-type semi-local functional
$E_{xc}^\text{in}$. The vdW-DF method proceeds to define a semi-local
and nonlocal functional components
\begin{align}
  E_{xc}^\text{vdW-DF} = E_{xc}^0 + E_c^\text{nl}\, ,
\end{align}
where the semi-local component satisfies
\begin{align}
  E_{xc}^0 \approx E_{xc}^{\rm in}
  \label{eq:semilocCond}
\end{align}
while the truly nonlocal XC energy term is defined as
\begin{align}
  E_c^\text{nl} = \int_0^\infty \, \frac{du}{2\pi}\,
  \text{Tr}\{\ln(\nabla \epsilon(iu) \cdot \nabla G) -
  \ln(\epsilon(iu))\}\, .
  \label{eq:ecnldef}
\end{align}

The vdW-DF framework can be interpreted as a rigorous implementation
of the Ashcroft picture of vdW forces since \eq{eq:ecnldef} formally
counts the shifts in electronic zero-point energies that arise with an
electrodynamical coupling between the internal GGA-type XC holes
\cite{HylBerSch14}. In the most widely used general-geometry versions,
the evaluation of \eq{eq:ecnldef} involves a second-order expansion
that allows an efficient universal kernel formulation
\cite{DioRydSch04, ThoCooLi07, BerIOP15}.  The vdW-DF versions are
entirely nonempirical and rest solely on the physics that underpins
the LDA XC energy and the GGA-type gradient-corrected exchange in
$E_{xc}^0$ and $E_{xc}^\text{in}$.

Here we benchmark the finite-temperature performance of the recent
consistent-exchange version vdW-DF-cx \cite{BerHyl14}. In this functional the 
exchange component in $E_{xc}^0$ is chosen to minimize
\begin{align}
  \delta E_{x}^0 \equiv E_{xc}^0 - E_{xc}^{\rm in},
  \label{eq:semilocCondmin}
\end{align}
for small-to-medium values of the scaled density gradient $s$. In practice, this means
that $\Delta E_{x}^0 = 0$ is for all systems but atoms and small molecules \cite{BerHyl14, HylBerSch14, BerIOP15} so that vdW-DF-cx effectively serves as an implementation of
(an expanded form of) the full vdW-DF framework \eq{eq:vdWDFfram} \cite{DioRydSch04,
  BerJCP14, HylBerSch14, BerIOP15}. Additional documentation for this new vdW-DF version 
  can be found in Refs.~\onlinecite{BerJCP14, HylBerSch14, BerIOP15}.

\section{Methodology}
\label{sect:methodology}

\subsection{Computational details}
\label{sect:computational-details}

DFT calculations were carried out using the projector augmented wave
(PAW) method \cite{Blo94, *KreJou99} as implemented in the Vienna
ab-initio simulation package (\textsc{vasp}) \cite{KreHaf93,
  *KreFur96a}. For vdW-DF-cx calculations we used the patch 
  released in Ref.~\onlinecite{torbjorn14}. In primitive 
  cell calculations the Brillouin zone was
sampled using $\Gamma$-centered $\vec{k}$-point grids with
$13\times13\times7$ divisions for hexagonal close packed (HCP)
structures, $14\times14\times14$ divisions for face-centered cubic
(FCC), and $15\times15\times15$ divisions for body-centered cubic
(BCC) structures. The plane wave cutoff energy was chosen 30\%\ larger
than the commonly recommended value for each element in order to
obtain very well converged forces and especially stresses. The values
employed are tabulated in Table~V of the Supplementary Material, which also provides details concerning the PAW setups.

\subsection{Vibrational modeling}

To evaluate finite temperature properties, we employed the quasi
harmonic approximation (QHA). First, the harmonic Helmholtz free
energy $F(T,V)$ was evaluated as a function of temperature at fixed
volume $V$ according to \cite{Wal98, phonopy}
\begin{align}
  F &= \frac{1}{2} \sum_{\vec{q}\nu} \hbar \omega_{\vec{q}\nu} + k_B T
  \sum_{\vec{q}\nu} \ln\left[ 1 - \exp(- \hbar \omega_{\vec{q}\nu} /
    k_B T) \right].
  \label{eq:helmholtz-free-energy}
\end{align}
Here, the summations are the result of a discretization of the
integral over the vibrational density of states and carried out over
phonon modes with momentum $\hbar\vec{q}$ and index $\nu$. The Gibbs
free energy $G(T,p)$ at constant pressure $p$ is obtained by repeating
the calculation of $F(T,V)$ for a range of volumes and minimizing the
sum of internal energy $U(V)$, Helmholtz energy $F(T,V)$ and the
pressure-volume term according to
\begin{align}
  G(T,p) &= \min_V\left[ U(V) + F(T,V) + pV \right].
  \label{eq:gibbs-free-energy}
\end{align}
While the internal energy $U(V)$, here, simply corresponds to the
Born-Oppenheimer energy as a function of volume, evaluation of the
vibrational contribution \eq{eq:helmholtz-free-energy} requires
knowledge of the phonon dispersion on a dense $\vec{q}$-point mesh. To
this end, force constants were calculated using the finite
displacement method and $4\times4\times4$ supercells. In the latter
calculations the Brillouin zone was sampled using $\Gamma$-centered
$3\times 3\times 3$ $\vec{k}$-point grids. The minimization in
\eq{eq:gibbs-free-energy} was carried out over volumes ranging from
$0.85 V_0$ to $1.15 V_0$, where $V_0$ is the volume corresponding to
the minimum of the Born-Oppenheimer energy landscape.

Knowledge of the Gibbs free energy as a function of volume and
temperature allows one to readily extract for example the lattice parameter(s), the bulk modulus and the thermal expansion
coefficient(s) at finite temperatures. All of these quantities were analyzed using the \textsc{phonopy} package \cite{phonopy}.

Furthermore, we calculated the cohesive energy $E_\text{coh}$ at zero Kelvin
including the zero-point energy (ZPE) contribution,
\begin{align}
  E_\text{coh} &= E_\text{bulk}
  + \frac{1}{2} \sum_{\vec{q}\nu} \hbar \omega_{\vec{q}\nu}
  - E_\text{atom},
\end{align}
where $E_\text{bulk}$ and $E_\text{atom}$ denote the total energy of
bulk material and atom, respectively. All terms in the latter equation
were evaluated at the 0\,K lattice constant corrected for zero-point
effects.

\subsection{Atomic reference energies: spin effects}
 
In general, spin-polarization must be included when calculating
$E_\text{atom}$. While a consistent spin-polarized version of the vdW-DF method 
was recently introduced \cite{ThoZulArt15} it has so far only been implemented in the \textsc{quantum-espresso} package \cite{Giannozzi09p395502}. The vdW-DF evaluation of 
nonlocal correlations amounts to tracking the total energy shift that arises with the electrodynamical coupling of plasmons, which, in turn, represent a GGA-type response to 
external fields \cite{HylBerSch14}. The vdW-DF approximations that are
implemented in \textsc{vasp} \cite{KliBowMic11} are not fully consistent
since they ignore the fact that spin polarization will itself 
adjust these plasmons \cite{ThoZulArt15}. 

In the present study we therefore proceeded as follows in order to obtain atomic reference energies and eventually cohesive energies for vdW-DF-cx. We calculated the non-spin-polarized atomic energy $E_\text{atom}^\text{nsp, vasp}$ using \textsc{vasp} and then added the atomic spin polarization energy $\Delta^\text{qe}_\text{spin}$ obtained using \textsc{quantum-espresso} with the rigorous-spin vdW-DF-cx description \cite{ThoZulArt15}. That is we obtained the atomic energies as
\begin{align}
	E_\text{atom} &= E_\text{atom}^\text{nsp, vasp}
	+ \underbrace{E_\text{atom}^\text{sp, qe} - E_\text{atom}^\text{nsp, qe}}_{\textstyle\Delta^\text{qe}_\text{spin}}.
 	\label{eq:spin-correct}
\end{align}
In effect this procedure amounts to computing \textsc{vasp} and QHA-based cohesive-energy estimates $E_\text{coh}^\text{vasp}$ and then adding a spin correction
\begin{align}
	\Delta_\text{spin-correction} = \Delta^\text{qe}_\text{spin} - 
	\Delta_\text{spin}^\text{vasp},
	\label{eq:delta-sp-corrdef}
\end{align}
where $\Delta_\text{spin}^\text{vasp} = E_\text{atom}^\text{sp, vasp} - E_\text{atom}^\text{nsp, qe}$ represents the \textsc{vasp} approximation for the vdW-DF-cx atomic spin polarization energy. A detailed compilation of the atomic reference energies can be found in Table~VI of the Supplementary Material.

In the \textsc{quantum espresso} calculations of $\Delta_\text{spin}$ we relied on norm-conserving pseudopotentials (NCPP) from the \textsc{abinit} package \cite{GonRigVer05}, using a plane-wave (density) cutoff of 80\,Ry (400\,Ry) so as to best mimic the fact the \textsc{vasp} calculations are based on hard PAW setups. This NCPP choice was possible for all but the case of W, where the \textsc{abinit} NCPP did not yield the correct spin polarization state. For the W case alone we therefore relied on a W ultrasoft pseudopotential in calculating $\Delta_\text{spin-correction}$.\footnote{We also tested using the W NCPP while constraining the spin polarization to the correct configuration, which yields a vdW-DF-cx value for the cohesive energy of W that is in even better agreement with experiment than when using the ultrasoft pseudopotential.}

\section{Results}
\label{sect:results}

\subsection{General assessment}

\begin{figure*}
  \centering
\includegraphics[width=0.9\linewidth]{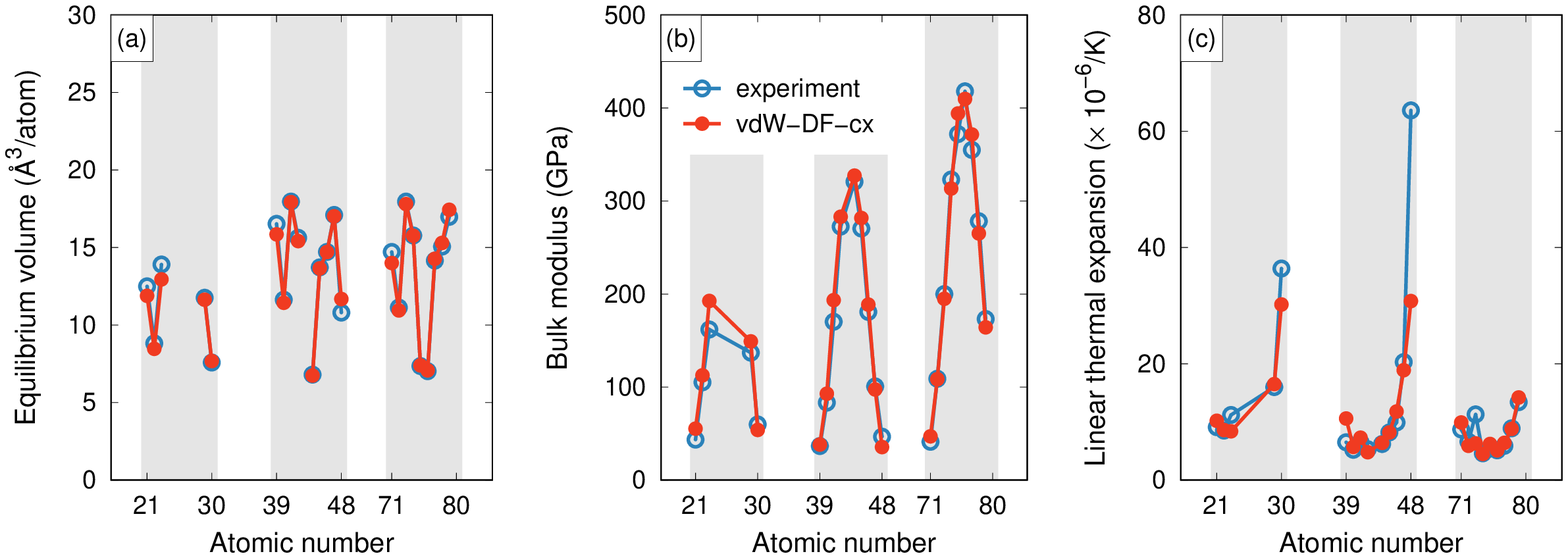}
\includegraphics[width=0.9\linewidth]{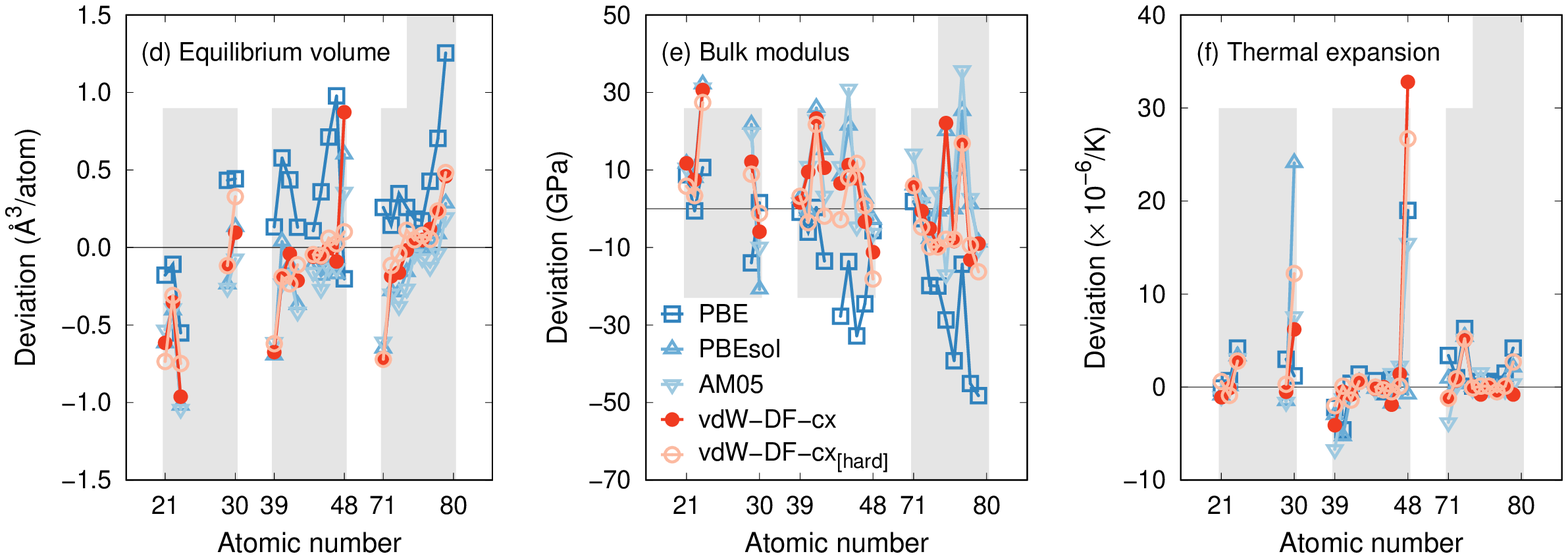}
  \caption{
    Overview of thermophysical properties at 300\,K obtained using the
    quasi-harmonic approximation in conjunction with density
    functional theory calculations. (a) Equilibrium volume, (b) bulk
    modulus, and (c) coefficient of average linear thermal expansion from
    experiment (Refs.~\onlinecite{KittelRefData,ThermalExpandData}) 
    and calculations based on the vdW-DF-cx functional. Deviation between different XC functionals and experiment for (d) equilibrium volume, (e) bulk modulus, and
    (f) coefficient of linear thermal expansion. The shaded regions
    indicate the set of $3d$, $4d$, and $5d$ transition metals.
  }
  \label{fig:comparison}
\end{figure*}

In the following, we provide an overview of the key results from our
comparative analysis of constraint-based XC functionals. A complete
compilation of the data obtained with each XC functional including
lattice constants can be found in the Supplementary Material.
To measure and compare the performance of different functionals, we consider the mean absolute percentage error (MAPE) defined as
\begin{align}
  M = \frac{1}{N} \sum_k
  \left|
  \frac{A_\text{DFT}^{(k)} - A_\text{expt}^{(k)}}{A_\text{expt}^{(k)}}
  \right|,
\end{align}
where $A_\text{DFT}^{(k)}$ and $A_\text{ref}^{(k)}$ denote predicted
and experimental values of a property of structure $k$ and the average
contains $N$ samples.

Many properties exhibit characteristic variations across the
transition metal series, which follow the $d$-band filling
[\fig{fig:comparison}] and are reproduced by all XC functional
considered here. While ZPE and thermal expansion effects are generally
limited to a few percent of the volume, they are nonetheless crucial
for an accurate assessment.

\begin{figure}
  \centering
\includegraphics[width=0.9\linewidth]{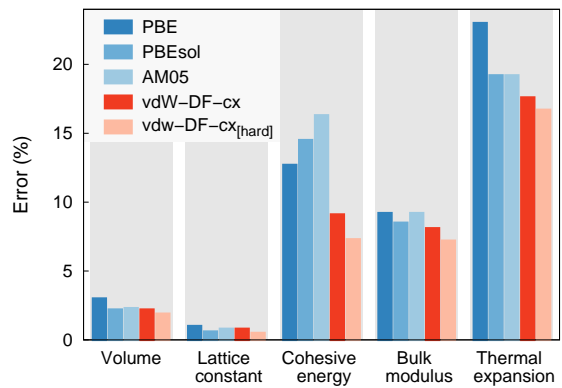}
  \caption{
    Performance of different constraint-based XC functionals based on the data from \tab{tab:performance-comparison}. The MAPEs were computed with regard to thermophysical properties measured at 300\,K, with the exception of the cohesive energy, for which zero K values are compared.
  }
  \label{fig:performance-comparison}
\end{figure}

\begin{table}
  \centering
  \caption{
    Performance of constrained based XC functionals with respect to
    the description of thermophysical properties.
    The comparison includes the equilibrium volume $V$, bulk modulus
    $B$, as well as the linear coefficient of thermal expansion
    $\alpha_l$ measured at room temperature, while in the case of the
    cohesive energy $E_{coh}$ zero Kelvin data are compared, albeit
    including ZPE effects.
    Unless otherwise noted the calculations were carried out using
    standard PAW setups.
    The comparison comprises 11 HCP, 7 FCC, and 5 BCC elements.
  }
\label{tab:performance-comparison}
  \newcommand{\scol}[1]{\multicolumn{1}{c}{#1}}
  \newcommand{\tcol}[1]{#1}
  \begin{tabularx}{0.9\columnwidth}{ll*{4}d}
    \hline\hline
    Functional
    & 
    & \scol{$V$}
    & \scol{$B$}
    & \scol{$\alpha_l$}
    & \scol{$E_{coh}$}
    \\
    \hline\hline

    \tcol{vdW-DF-cx}
    & HCP	 & 3.0\% & 9.9\% & 20.6\% & 10.9\% \\
    & FCC	 & 1.0\% & 5.0\% &  5.9\% & 9.5\% \\
    & BCC	 & 1.9\% & 8.4\% & 27.3\% & 4.5\% \\
    & total  & 2.2\% & 8.1\% & 17.6\% & 9.1\% \\[6pt]
    
    \tcol{vdW-DF-cx}
    & HCP	 & 2.6\% &  8.5\% & 17.7\% & 7.8\% \\
    \tcol{\phantom{x}(hard PAW}
    & FCC	 & 1.0\% &  4.9\% &  5.5\% &  8.7\% \\
    \tcol{\phantom{x}setups)}
    & BCC	 & 1.7\% &  7.7\% & 30.0\% &  4.2\% \\
    & total  & 1.9\% &  7.2\% & 16.7\% &  7.3\% \\[6pt]
    
    \tcol{PBE}	                           
    & HCP	 & 2.3\% &  7.1\% & 23.6\% & 10.9\% \\
    & FCC	 & 4.6\% & 15.1\% & 11.8\% &  8.1\% \\
    & BCC	 & 2.2\% &  5.6\% & 37.4\% & 23.3\% \\
    & total  & 3.0\% &  9.2\% & 23.0\% & 12.7\% \\[6pt]
    
    \tcol{PBEsol}    		              
    & HCP	 & 2.8\% &  9.9\% & 21.7\% & 14.7\% \\
    & FCC	 & 1.1\% &  6.1\% &  8.4\% & 12.4\% \\
    & BCC	 & 2.6\% &  9.0\% & 28.7\% & 16.8\% \\
    & total  & 2.2\% &  8.5\% & 19.2\% & 14.5\% \\[6pt]
    
    \tcol{AM05}	                           
    & HCP	 & 2.7\% & 11.3\% & 22.3\% & 25.7\% \\
    & FCC	 & 1.2\% &  6.8\% &  6.8\% &  9.7\% \\
    & BCC	 & 3.0\% &  7.8\% & 29.8\% &  4.8\% \\
    & total  & 2.3\% &  9.2\% & 19.2\% & 16.3\% \\
    
    \hline\hline
  \end{tabularx}
\end{table}

The performance comparison [\fig{fig:performance-comparison} 
and \tab{tab:performance-comparison}] confirms that PBEsol and, 
with the exception of the cohesive energy, also AM05 represent 
general improvements over PBE. The relatively large MAPEs
[\tab{tab:performance-comparison}] arise mostly from larger 
errors in just a few systems. In the case of PBE the MAPE for 
the cohesive energy of BCC structures is particularly large. 
This issue is primarily caused by an inaccurate description 
of the electronic configuration of the isolated spin-polarized 
atoms, which impacts the atomic reference energy.

More interestingly, the comparison demonstrates
that the truly nonlocal vdW-DF-cx performs at least at the level of 
PBEsol and AM05. In fact, considering all properties
vdW-DF-cx provides the best overall agreement with the experimental
reference data. This is especially remarkable since previous
non-empirical versions of the DF method, namely vdW-DF1 \cite{DioRydSch04} 
(in which exchange is approximated by the revPBE functional\cite{ZhYa98}) 
and vdW-DF2 \cite{LeeMurKon10} (in which exchange is approximated by a 
revised version \cite{MurLeeLan09} of the PW86 functional \cite{PerWan86}) perform 
rather poorly for the late transition metals. In particular, in the case of 
Ag and Au the lattice constants are considerably overestimated in vdW-DF1 and
vdW-DF2,\cite{KliBowMic11} while vdW-DF-cx yields excellent results for these elements.

\subsection{Cohesive energies}

\begin{figure}
  \centering
\includegraphics[width=0.9\linewidth]{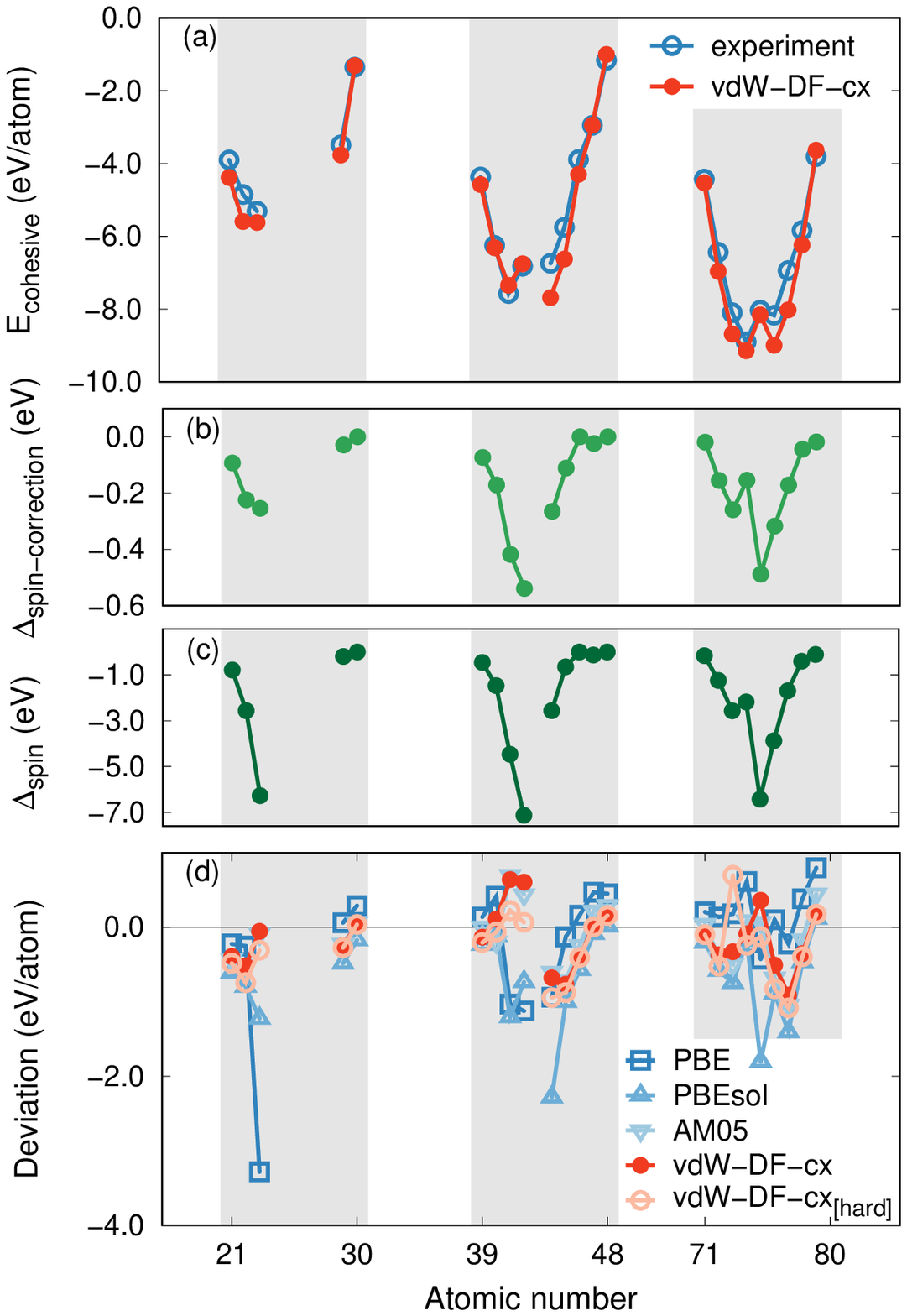}
  \caption{
    (a) Cohesive energies at 0\,K obtained using the
    quasi-harmonic approximation in conjunction with density
    functional theory calculations from experiment
    (Ref.~\onlinecite{KittelRefData}) and calculations based on the
    vdW-DF-cx functional including the spin correction according to \eq{eq:delta-sp-corrdef}.
    (b) Spin correction according to \eq{eq:delta-sp-corrdef} that is
    added to the \textsc{vasp}-based description for our vdW-DF-cx
    benchmark to account for a rigorous description of spin
    effects in the atomic reference energies. (c) Atomic spin
    polarization energies $\Delta_\text{spin}$ defined in
    \eq{eq:spin-correct} as calculated in the spin vdW-DF-cx
    formulation (available via \textsc{quantum-espresso}
    calculations).  (d) Deviation between different XC functionals and
    experiment for cohesive energies. The shaded regions indicate the
    set of $3d$, $4d$, and $5d$ transition metals.  }
  \label{fig:Ecoh-comparison}
\end{figure}

Overall the constraint-based functionals considered here perform
reasonably well with regard to the description of the cohesive energy
[\fig{fig:Ecoh-comparison}] although for most of the functionals there
are problems with specific elements. Most notably, the vdW-DF-cx
description clearly outperforms the other functionals in terms of the
cohesive energies.

Moreover, the results demonstrate that the rigorous
inclusion of spin effects in vdW-DF-cx\cite{ThoZulArt15} is important
for an accurate description of the cohesive energy in nonmagnetic
transition metals [\fig{fig:Ecoh-comparison}(b)]. Since the atomic
spin polarization energies are very large in the middle of the
transition-metal bands [\fig{fig:Ecoh-comparison}(c)], it is important
to use the rigorous-spin vdW-DF-cx formulation \cite{ThoZulArt15}. The
corrections are negative and systematically lead to larger values for
$\Delta_\text{spin}$ (see Table~VI of the Supplementary Material). As
a consequence our rigorous-spin vdW-DF-cx calculations provide a
systematic improvement for the description of nonmagnetic transition
metals, lowering for example the MAPE from 9.4\% to 7.3\%\ when using
hard PAW setups for vdW-DF-cx.

\subsection{Effect of PAW setups in the case of vdW-DF-cx}

For computational efficiency it is often desirable to employ PAW
setups that contain only the highest occupied states in the
valence. This not only limits the total number of states in the
calculation but also often allows using relatively large core radii
that require smaller plane wave basis cutoffs in order to obtain
converged results. While so far we have only considered results
obtained using such ``standard'' PAW setups, it is now instructive to
examine the choice of the PAW setup more closely. To this end, we
exclusively consider calculations based on the vdW-DF-cx functional
and ``hard'' PAW setups as detailed in Table~V of the Supplementary Material.

Using the hard PAW setups systematically improves the agreement with
experiment, typically reducing the MAPE by a fraction of a percent
(\tab{tab:performance-comparison} and
\fig{fig:performance-comparison}). Yet, the comparison clearly
demonstrates that already the ``standard'' PAW setups yield very good
result and are sufficient to achieve good results in many situations.

Cadmium represents an exception, for which there is pronounced
difference between standard and hard setups. For example, the lattice
constants at 300\,K change from $a=3.168\,\text{\AA}$ and
$c=5.373\,\text{\AA}$ to $a=3.023\,\text{\AA}$ and
$c=5.512\,\text{\AA}$ when going from standard to hard PAW setups. The
latter values are also in notably better agreement with the
experimental numbers of $a=2.98\,\text{\AA}$ and $c=5.62\,\text{\AA}$
(Table~II of the Supplementary Material). More generally, the late transition
metals in HCP structure (Zn and Cd) are challenging for all XC
functionals. This behavior is related to their special electronic
structure that manifests itself e.g., in $c/a$-ratios (experimentally
$c/a=1.89$ and 1.86 for Cd and Zn, respectively) that are considerably
larger than in the ideal HCP structure ($c/a=1.633$).

\subsection{Structure trends, semi-local and nonlocal functionals }

The deviations between calculated and experimental data follow certain
trends [\fig{fig:comparison}(d-f)]. While PBE tends to overestimate
the equilibrium volume, the other functionals are overall in rather
close agreement with the reference data.

\begin{figure}
  \centering
\includegraphics[width=0.9\linewidth]{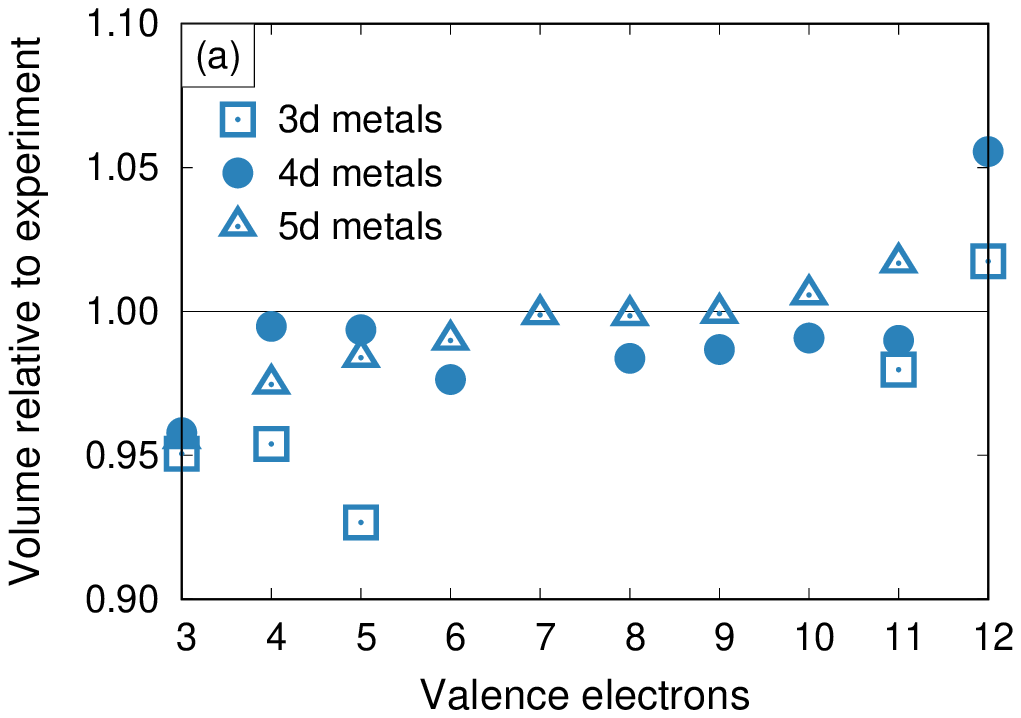}
\includegraphics[width=0.9\linewidth]{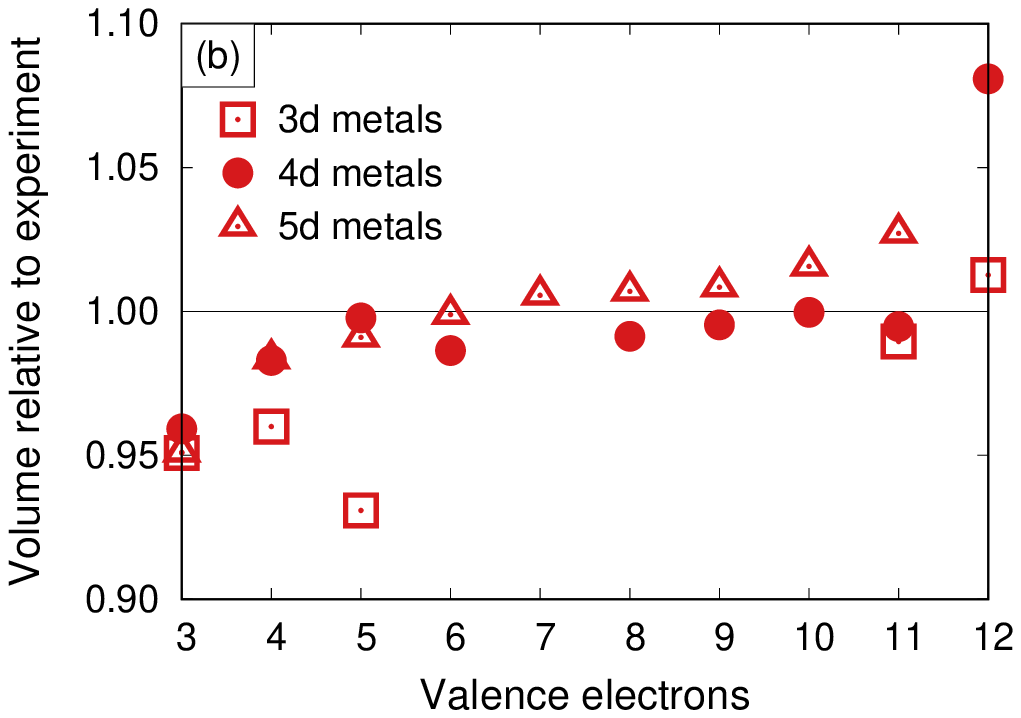}
  \caption{
    Ratio of calculated to experimental equilibrium volumes for (a)
    PBEsol and (b) vdW-DF-cx (hard PAW setups). The transition from a
    volume underestimation converting to a volume overestimation is
    expected for GGAs and is found in both cases.
  }
  \label{fig:volume-comparison}
\end{figure}

Contrasting specifically vdW-DF-cx and PBEsol performance
(\fig{fig:volume-comparison}) shows that both functionals exhibit
similar trends with respect to the variation of the accuracy with
$d$-band filling. It is apparent that the data for the first and last
columns of the series are slightly under and overestimated,
respectively. The largest relative corrections of the volume arise for
Zn and Cd and those elements also have some of the largest vibrational
corrections to the cohesive energies. In fact, most DFs provide an inaccurate description of these elements, which as indicated
above exhibit a HCP structure with a very large axial ratio.

Larger deviations from the reference data are also observed for V
(BCC) for all functionals. We ascribed this behavior to the low
temperature magnetism that has been reported in this element
\cite{SmiFin66}, while in the present calculations it is treated
without spin polarization.

Similar trends as for the equilibrium volume can be observed for bulk
modulus [in reversed fashion, \fig{fig:comparison}(e)] and linear
coefficient of thermal expansion [\fig{fig:comparison}(f)] although
the errors are more scattered. The latter effect is probably connected
to a larger uncertainty in the experimental data as will be discussed
in the next section.

\subsection{Bulk modulus and thermal expansion}
\label{sect:bulk-modulus}

So far we used experimental values from compilations of standard
values \cite{KittelRefData, ThermalExpandData} as reference data for
equilibrium volumes (lattice constants), bulk moduli, and thermal
expansion coefficients. While the data for lattice constants are
usually very accurate, it must be acknowledged that measurements of
bulk moduli and thermal expansion coefficients can carry rather
significant errors, which are usually not documented in reference
compendia. A closer inspection of experimental data available in the
general scientific literature, however, reveals that at least in some
cases these errors can be comparable or even exceed the deviation
between the best performing XC functionals and experiment.

\begin{figure}
  \centering
\includegraphics[width=0.9\linewidth]{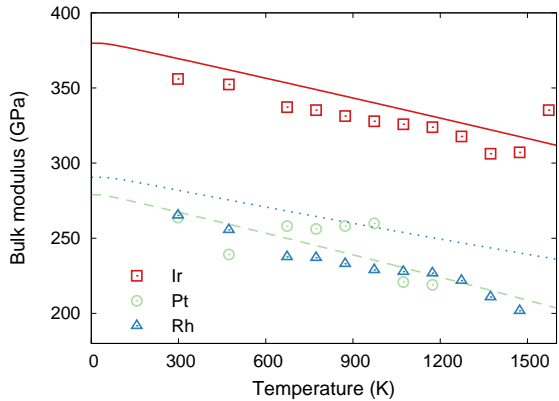}
  \caption{
    Temperature dependence of bulk modulus from experiment and
    calculation (vdW-DF-cx). The experimental data points were
    obtained by converting the data for Young's modulus $E$ and
    Poisson ration $\nu$ measured in Ref.~\onlinecite{MerLupTop01}
    using the relation $B=E/3(1-\nu)$.
  }
  \label{fig:experimental-bulkmoduli}
\end{figure}

For illustration, we employ experimental data for the Young's moduli
$E$ and Poisson ratios $\nu$ of polycrystalline samples of the FCC
metals Ir, Pt, and Rh \cite{MerLupTop01}. As the experimental data
range from 300 to 1500\,K, this also allows us to compare experiment
and calculations over a wide temperature span. The experimental data
can be converted to bulk modulus using the relation
$B=E/3(1-\nu)$. This illustrates that there is considerable scatter in
the experimental data with changing temperature that does not appear
to be associated with a specific trend
(\fig{fig:experimental-bulkmoduli}); this is particularly pronounced
in the case of Pt. The calculations overestimate the experimental data
but overall the agreement is good with similar temperature dependence.

\subsection{General discussion}

High accuracy and transferability of vdW-DF-cx had been previously
indicated by a range of successful applications to systems that
combine regions of both sparse and dense electron distributions
\cite{BerHyl14, ErhHylLin15, RanPRB16, BrownAltvPRB16, BerJCP14,
  ThoZulArt15, Arnau16, LofGroMot16}. In the present paper it has
been demonstrated that, unlike the vdW-DF1 and vdW-DF2 versions, vdW-DF-cx also
performs very well for hard materials.

The strong performance of PBEsol for traditional materials can be
primarily traced to two of its features, namely a good form for
gradient-corrected exchange and a good balance between this exchange
part and the account of gradient-corrected correlation.  The vdW-DF-cx
strategy of seeking an ACF evaluation, see \eq{eq:vdWDFfram}, implies
picking a semi-local exchange component in $E_{xc}^0$ that is given by
diagramatic expansion and therefore is similar to 
that of PBEsol in the low-to-medium $s$ regime. Yet, vdW-DF-cx still
replaces the PBEsol description of gradient-corrected correlation
entirely with a truly nonlocal XC term $E_c^\text{nl}$. The fact that
DF-cx performs at a PBEsol level with respect to hard materials thus
suggests that vdW-DF-cx achieves a good balance between exchange and
correlation. This observation makes it plausible that one can obtain
further functional improvements by relying on the ACF recast, 
\eq{eq:vdWDFfram}, for descriptions of nonlocal correlation effects
\cite{DioRydSch04, BerHyl14, HylBerSch14, BerIOP15}.

\section{Summary}
\label{sect:summary}

This study presents a comprehensive benchmark of constraint-based
semi-local and non-local functionals with respect to finite
temperature thermophysical properties of nonmagnetic transition
metals. The main outcome of this comparison is that, unlike its
predecessors in the vdW-DF family, the recently developed non-local
vdW-DF-cx version achives good transferability and accuracy also for
hard materials. This is crucial, for example, for investigations of
weakly-bound molecules at transition-metal surfaces. In the case of
vdW-DF1 and vdW-DF2, the substantial overestimation of the lattice
constants of the late transition metals, in particular Ag and Au,
limited their application to these systems. The successful validation
of vdW-DF-cx for these cases allows full ionic relaxation and thus
tracking the impact of e.g., associated adsorption-induced surface
modifications \cite{KliBowMic11, BerJCP14, Arnau16}.

In completing our work we found that Ambrosetti and Silvestrelli
\cite{AmbSil16} recently presented a related benchmarking of how a
range of functionals, including vdW-DF-cx, performs for the coinage
metals. Our findings are consistent with their results but we have a
broader scope as we consider the entire set of nonmagnetic transition
metals and assess performance by comparing with experiments on a range
of thermo-physical properties. This broader scope is instrumental for
asserting the transferability of the method and provides a strong test
of the robustness of the underlying physical input.

For the other functionals considered here the excellent performance is
expected but it is interesting to note that the truly nonlocal
functional vdW-DF-cx has as good if not better performance and
transferability. This is encouraging for further development that
builds of the vdW-DF framework.

Finally, we observe that quantitative comparisons as in the tables
included in the Supplementary Material can also assist a crude
benchmarking of future vDW-DF versions. This is because the tables
implicitly provide a quantification of the net differences between the
raw Kohn-Sham results and the associated room temperature
characterizations that are relevant for comparison with most
experimental observations.

\begin{acknowledgments}
We thank Kristian Berland for discussions. This work has been
supported by the Swedish Research council (VR) and the Knut and Alice
Wallenberg foundation as well as computer time allocations by the
Swedish National Infrastructure for Computing at NSC (Link\"oping),
HPC2N (Ume{\aa}), and C3SE (Gothenburg).
\end{acknowledgments}

\end{document}